\documentclass[
prx
,twocolumn
,nofootinbib
,floatfix
,superscriptaddress
]{revtex4-2} 

\usepackage{notes2bib}
\usepackage{natbib}
\usepackage{gensymb}
\usepackage{amsthm}
\usepackage{amsmath}
\usepackage{color}
\usepackage{bm}
\usepackage{amsfonts}
\usepackage{dsfont}
\usepackage{graphicx}
\usepackage{float}
\usepackage{amsfonts}
\usepackage[justification=justified,singlelinecheck=false]{caption}
\usepackage{subcaption}
\usepackage{verbatim}
\usepackage{amsfonts}
\usepackage{bbold}
\usepackage{dcolumn}
\usepackage{bm}
\usepackage[dvipsnames]{xcolor}
\usepackage{listings}
\definecolor{blueprl}{RGB}{46,48,146}
\usepackage[pdftex,colorlinks=true,urlcolor=blueprl,citecolor=blueprl,linkcolor=blueprl]{hyperref}
\usepackage{mathrsfs}
\usepackage{mathtools}
\usepackage{times}
\usepackage{color}
\usepackage[dvipsnames]{xcolor}
\usepackage{braket}

\usepackage{soul}

\newcommand{\nonvt}{\vphantom{\frac{1}{\sqrt{2}}} \nonumber} 
\newcommand{\mrm}[1]{\mathrm{#1}}

\usepackage{mathtools}
\usepackage{dsfont}
\usepackage[mathscr]{euscript}
\usepackage{amsmath}
\usepackage{amssymb}
\usepackage{graphicx}
\usepackage{dcolumn}
\usepackage{bm}
\usepackage{epsfig}
\usepackage{latexsym,mathrsfs}
\usepackage{graphicx}
\usepackage{caption}
\usepackage{color}
\usepackage{hyperref}
\usepackage{float}
\usepackage{diagbox}
  \usepackage{cancel}
  \usepackage[normalem]{ulem}

\definecolor{vividviolet}{rgb}{0.62, 0.0, 1.0}
\definecolor{amaranth}{rgb}{0.9, 0.17, 0.31}
\definecolor{palatinateblue}{rgb}{0.15, 0.23, 0.89}
\definecolor{brightpink}{rgb}{1.0, 0.0, 0.5}
\definecolor{cornflowerblue}{rgb}{0.39, 0.58, 0.93}
\definecolor{deepcarminepink}{rgb}{0.94, 0.19, 0.22}
\definecolor{radicalred}{rgb}{1.0, 0.21, 0.37}
\definecolor{blueblue}{RGB}{21,47,181}
\definecolor{greengreen}{RGB}{65,166,16}

\usepackage{xtab,afterpage}
\usepackage{hyperref}
\usepackage{setspace,calc}

\newcommand{\vt}{\vphantom{\frac{1}{2}}}

\usepackage[pdftex,colorlinks=true,urlcolor=blueprl,citecolor=blueprl,linkcolor=blueprl]{hyperref}

\newcommand{\be}{\begin{equation}}
\newcommand{\ee}{\end{equation}}
\newcommand{\bs}{\begin{split}} 
\newcommand{\bea}{\begin{eqnarray}}
\newcommand{\eea}{\end{eqnarray}}
\newcommand{\infint}{\int_{-\infty }^{\infty }}

\newcommand{\D}{\mathrm{d}}
\newsavebox{\myhbar}

\allowdisplaybreaks

\begin{document}

\newcommand{\CQCT}{\affiliation{Centre for Quantum Computation \& Communication Technology, School of Mathematics \& Physics, The University of Queensland, St.~Lucia, Queensland, 4072, Australia}}
\newcommand{\FUB}{\affiliation{Dahlem Center for Complex Quantum Systems, Freie Universit\"at Berlin, 14195 Berlin, Germany}}
\newcommand{\WAT}{\affiliation{Department of Physics and Astronomy, University of Waterloo, Waterloo, Ontario, Canada, N2L 3G1}}
\newcommand{\PI}{\affiliation{Perimeter Institute, 31 Caroline St., Waterloo, Ontario, N2L 2Y5, Canada}}
\newcommand{\EQUS}{\affiliation{Centre for Engineered Quantum Systems, School of Mathematics and Physics, The University of Queensland, St. Lucia, Queensland, 4072, Australia}}
\newcommand{\SU}{\affiliation{Department of Physics, Stockholm University, AlbaNova University Center, SE-106 91 Stockholm, Sweden}}
\newcommand{\SIT}{\affiliation{Department of Physics, Stevens Institute of Technology, Castle Point Terrace, Hoboken, New Jersey 07030, U.S.A.}} 

\title{Signatures of Rotating Black Holes in Quantum Superposition}
\author{Cendikiawan Suryaatmadja}
\WAT 

\author{Cemile Senem Arabaci}
\WAT 

\author{Matthew P.\ G.\ Robbins}
\WAT 

\author{Joshua Foo}
\CQCT
\SIT 

\author{Magdalena Zych}
\EQUS 
\SU 

\author{Robert B.\ Mann}
\WAT 
\PI 

\begin{abstract}
A new approach for operationally studying the effects of spacetime in quantum superpositions of semiclassical states has recently been proposed by some of the authors.
This approach was applied to the case of a (2+1)-dimensional Ba\~nados-Teitelboim-Zanelli (BTZ) black hole in a superposition of masses, where it was shown that a two-level system interacting with a quantum field residing in the spacetime exhibits resonant peaks in its response at certain values of the superposed masses. Here, we extend this analysis to a mass-superposed \textit{rotating} BTZ black hole, considering the case where the two-level system co-rotates with the black hole in a superposition of trajectories. We find similar resonances in the detector response function at rational ratios of the superposed outer horizon radii, specifically in the case where the ratio of the inner and outer horizons is fixed. This suggests a connection with Bekenstein's seminal conjecture concerning the discrete horizon spectra of black holes in quantum gravity, generalized to the case of rotating black holes. Our results suggest that deeper insights into quantum-gravitational phenomena may be accessible via tools in relativistic quantum information and curved spacetime quantum field theory. 
\end{abstract}

\date{\today}

\maketitle

\section{Introduction}
The study of black holes has captivated physicists for the past century, with the goal of unraveling the rich tapestry of phenomena they manifest. The early work of Bekenstein laid the foundation for our modern understanding of black hole thermodynamics, linking the properties of black holes with principles of statistical mechanics. In his seminal paper \cite{PhysRevD.7.2333}, Bekenstein conjectured that black holes possess an entropy proportional to the area of their event horizon. This theory was later substantiated by Hawking's calculation of black hole radiation, implying a profound connection between gravity, thermodynamics and quantum mechanics \cite{hawking1974black}. 

In addition to his work on black hole entropy, Bekenstein put forth a conjecture concerning the discreteness of black hole mass. Bekenstein's hypothesis posited that black hole horizon $r_H$ should possess a discrete spectrum, akin to the quantized energy levels of atoms \cite{Bekenstein:2020mmb}, i.e.\ 
\begin{align}
    r_H &= n \hslash , \:\:\: n = 1, 2 , \hdots 
\end{align}
Bekenstein's analogical treatment of black holes as atomic systems motivates the possibility that we can consider them in superpositions of masses \cite{dipumpoPRXQuantum.2.040333}, each corresponding to a different spacetime, unique up to a general coordinate transformation.

Recently a method for studying the signatures of spacetimes in quantum-controlled superpositions of different configurations
was proposed by several of the authors of this work \cite{Foo_2021desitterschrodingercat}.
This approach, which is grounded in the tools of quantum field theory in curved spacetime, considers the evolution of a two-level system coupled to the quantum spacetime through its interaction with an ambient quantum field. The response of such a two-level system (i.e.\ a quantum detector \cite{perchePhysRevD.101.045017,brownPhysRevD.87.084062,Louko_2008,Louko_2006}) situated at a fixed radius outside a mass-superposed nonrotating (2+1)-dimensional Ba\~nados-Teitelboim-Zanelli (BTZ) black hole \cite{BTZPhysRevLett.69.1849} was specifically considered in Ref.~\cite{Foo_2022}, and shown  to experience resonant behaviour when the ratio of 
the horizon radii (each proportional to the the square root of the mass) had a rational value, in accord with Bekenstein's mass discretization conjecture. This approach was applied also to a de Sitter spacetime in a superposition of cosmological constants \cite{Foo_2021desitterschrodingercat} and Minkowski spacetime in a superposition of periodic boundary conditions \cite{foo_topologiesPhysRevD.107.045014}.  It  represents a unique perspective in understanding the phenomenology of spacetime superpositions \cite{dewitt2011role,Anastopoulos_2015,akilsemiclassicalPhysRevD.108.044051}, which has been considered in the context of low-energy quantum gravity \cite{Belenchia:2018szb,CHRISTODOULOU201964,Chen2023quantumstatesof,Galley2022nogotheoremnatureof}, generalizations of quantum reference frames
\cite{delaHamette2023,giacomini10.1116/5.0070018,hoehn2023matter}, decoherence \cite{Arrasmith2019,kieferPhysRevD.53.7050}, and indefinite causal structure \cite{Bell4time, Henderson:2020zax}.

In this paper, we generalize the prior work of Foo et.~al. \cite{Foo_2022}
by considering BTZ black holes with non-zero angular momenta. Incorporating non-zero angular momenta into the BTZ black hole framework introduces new challenges and complexities. For one, it is no longer computationally tractable to consider detector trajectories that are static outside the black hole; the most convenient choice involves a superposition of co-rotating frames of the detector. Moreover, the addition of rotation alters the geometry of the black hole, leading to the formation of an inner and outer event horizon which raises the question whether resonances analogous to those demonstrated in \cite{Foo_2022} at all manifest for rotating black holes. 

Indeed, our main result is the demonstration of similar resonances in the detector response at discrete values of the rotating black hole's outer horizon radius. This in turn provides indirect corroboration of Bekenstein's conjecture applied to rotating BTZ black holes. In particular, resonances (i.e sharp increases in transition probability) occur at rational values of the ratio between the outer horizons of the superposed black hole, provided the ratios of the outer to inner horizon radii are the same for each amplitude of the superposition. Our findings provide further evidence that insights into quantum-gravitational phenomena, in particular those arising from quantum superpositions of geometries, are accessible via tools in relativistic quantum information and curved spacetime quantum field theory.

The outline of our paper is as follows. In Sec.\ \ref{sec:BTZ}, we review the geometric and field-theoretic background of the rotating BTZ spacetime and introduce the quantum-controlled Unruh-deWitt model. In Sec.\ \ref{sec:Equal angular momentum}, we present our results concerning the response of a detector co-rotating with superposed black holes of equal angular momentum and in Sec.\ \ref{sec:fixed ratio} we explore the case where the inner to outer horizon ratio of each superposed black hole are treated as fixed parameters.  In particular, we draw a connection between the detector response (i.e.\ the peaks in its response function at rational values of the outer horizon radii) and Bekenstein's conjecture (i.e.\ that quantum black holes have a discrete horizon spectrum, which for rotating BTZ black holes, corresponds to integer values of the outer horizon radius regarding the discrete horizon spectrum for quantum black holes \cite{Kwon_2010}). We conclude with some final remarks in Sec.\ \ref{sec:conclusion}. We will henceforth use the natural units, $ \hbar = k_B = c = 8G = 1$.

\section{Rotating BTZ Black Hole}\label{sec:BTZ}

\subsection{Geometry}

The rotating BTZ black hole is a (2+1) dimensional solution to Einstein's field equations in anti-de Sitter spacetime with cosmological constant $\Lambda= - 1/l^2$, with $l$ being the AdS length scale. The general form of the metric is 
 \cite{BTZPhysRevLett.69.1849}
\begin{align}
    \D s^2 &=\frac{(r^2-r_+^2)(r^2-r_-^2)}{l^2r^2} \D t^2-\frac{l^2r^2}{(r^2-r_+^2)(r^2-r_-^2)} \D r^2 
    \nonvt 
    \\
    & \qquad - r^2\left( \D \phi-\frac{r_+r_-}{lr^2} \D t\right)^2\label{metric}
\end{align}
and can be obtained from an angular ``boost'' of the static BTZ black hole, followed by a re-identification of the coordinates 
\cite{PhysRevD.86.064031,Lemos:1995cm,Hennigar:2020drx}. Here
 $(t,r,\phi)$ represent the polar space-time coordinate with origin at the center of the black hole.
The angular coordinate $\phi \in [0,2\pi]$, and  
 $r_-$ and $r_+$ are respectively the inner and outer horizon of the black hole, which for a BTZ black hole of mass $M$ and angular momentum $J$ are defined by the relations,
\begin{align}
    M &=\frac{r_+^2+r_-^2}{l^2}  
    \\
    J &=\frac{2r_+r_-}{l}\label{relation} 
\end{align}
where the inner $(r_-)$ and outer $(r_+)$ horizons have the following dependence on $M,J,l$:
\begin{align}
    r_\pm &=\frac{\sqrt{l^2M\pm l\sqrt{l^2M^2-J^2}}}{\sqrt{2}} \label{explicitrplus} 
\end{align}

\subsection{Scalar Field Quantization}
In our analysis, we are interested in the quantization of a scalar field on the black hole spacetime, when its mass and angular momentum is in a superposition of different values. Let us first consider a conformally-coupled scalar field (in the AdS vacuum). For the BTZ spacetime, we can write the two-point correlation function, $W(x,x') \equiv \langle 0 | \hat{\phi}(x) \hat{\phi}(x') | 0 \rangle$ (also called the Wightman function), as a sum of images of vacuum Wightman functions in AdS$_3$ spacetime \cite{PhysRevD.49.1929, Carlip_1998}:
\begin{align}
    W_\mrm{BTZ}(x,x')= \frac{1}{\mathcal{N}} \sum_{n,m=-\infty}^\infty \eta^nW_{\mathrm{AdS}_3}(\Gamma^n x,\Gamma^mx')
    \label{eq4}
\end{align}
Here, $\eta=\pm1$ corresponds to the twisted ($\eta=-1$) or untwisted ($\eta=1$) nature of the field (for simplicity, we consider only the $\eta = +1$ case here) and $\Gamma: (t,r,\phi)\to(t,r,2\pi n )$ is a discrete isometry enacted on the coordinates. $W_{\mathrm{AdS}_3}$ is the vacuum Wightman function in (2+1)-dimension. In terms of the field itself, this identification corresponds to the transformation, 
\begin{align} 
    \hat{\phi}(x):=\frac{1}{\sqrt{\mathcal{N}}}\sum_{n
    = - \infty}^{+\infty} \eta^n\hat{\psi}(\Gamma^nx)
\end{align}
where $\hat{\psi}$ is an unidentified massless scalar field. The normalization term 
$\mathcal{N}=\sum_n \eta^{2n}$ preserves the canonical commutation relation,  $[ \phi(x) , \phi(x')] = \delta(x - x') + \mathrm{image \: terms}$. An advantage of studying the BTZ black hole spacetime is that the Wightman functions are relatively straightforward to compute (only requiring an image sum, rather than a mode sum). 

\subsection{Quantum-Controlled Unruh-deWitt Model}

\noindent Let us now introduce the operational framework {utilized in Refs.\ \cite{Foo_2022,foo_topologiesPhysRevD.107.045014} that allows us to study the phenomenological effects produced by the rotating spacetime in superposition. In our analysis, we adopt standard postulates about the superposition of spacetime metrics. In particular, We assume that:
\begin{enumerate}
    \item The gravitational field sourced by a massive object can exist in a superposition of ``coherent states,'' each amplitude giving rise to a semiclassical metric $g$. 
    \item The dynamics of systems residing in the spacetime undergo the usual time evolution, as governed by quantum mechanics, on the respective amplitudes of the superposition.
\end{enumerate}
In particular, we consider the spacetime produced by a rotating BTZ black hole in a symmetric superposition of mass and angular momenta $(M_A, J_A), (M_B,J_B)$. We assume, as in earlier studies  \cite{Foo_2022}, that the field is in the AdS vacuum $| 0 \rangle$ and therefore factors from the state of the metric. Our model is not restricted to such a simplification; one can generally consider field states that depend on the spacetime. Finally, we consider a two-level system, modelled here as an Unruh-deWitt detector initially in its ground state $| g \rangle$, coupled to the field and the metric. The Hilbert space of the entire system can thus be written as the product $\mathcal{H}=\mathcal{H}_\mathrm{BH}\otimes\mathcal{H}_\mathrm{F}\otimes\mathcal{H}_\mathrm{D}$ where $\mathcal{H}_\mathrm{BH}$, $\mathcal{H}_\mathrm{F}$, $\mathcal{H}_\mathrm{D}$ are respectively the Hilbert space of the black hole, massless field and the detector. With this in mind, the initial state $| \psi \rangle \equiv | \psi(t_i) \rangle$ can be written as, 
\begin{align}
    | \psi \rangle &= \frac{1}{\sqrt{2}} ( | M_A , J_A \rangle + | M_B , J_B \rangle ) \otimes | 0 \rangle \otimes | g \rangle . 
\end{align}
We study the superposed rotating black hole under two parametric constraints. Initially we look at superpositions with equal angular momentum, $J_A = J_B$. We find that the resonant behaviour present in the static case rapidly diminishes as the angular momentum increases.  We then consider fixed values of the ratio of the outer to inner horizon radii for each black hole state. We find that the resonant behaviour mentioned previously is robust against changes in $J$ provided these ratios are the same for each state in the superposition. 

The interaction between the black hole, quantum field, and detector is modelled using the interaction Hamiltonian,
\begin{align}
& H_\mathrm{int} (\tau) \label{eq9}\\
 & \quad =  \lambda \sum_{D \:= \: A,B} 
  \chi_D(\tau)\hat{\mu}_D(\tau) \hat{\phi}(x_D)\otimes \ket{M_D,J_D}\bra{M_D,J_D} ,  \nonumber
\end{align}
where $\lambda\ll1$ is the interaction strength, $\tau$ is the proper time corresponding to the BTZ spacetime given by the black hole of mass $M_D$ and angular momentum $J_D$, $\chi_D(\tau)$ is the switching function describing how the detector turns  its interaction with the field on and off, $\mu_D(\tau)=\ket{e_D}\bra{g_D}e^{i\Omega_D\tau}+\ket{g_D}\bra{e_D}e^{-i\Omega_D\tau}$ is the monopole moment between  ground and excited states, and $\hat{\phi}(x_D)$ is the massless field operator parametrized by coordinates $x_D\equiv x_D (\tau)$ of the detector's worldline. From Eq.\ (\ref{eq9}), we note that the structure of the interaction Hamiltonian takes an identical form in either branch of the spacetime superposition; the control $| M_D, J_D \rangle\langle M_D ,J_D| $ determines ``which spacetime'' the detector couples to. Throughout this paper, we consider that the internal structure of the detector as well as its interaction with the field are independent of the black hole mass, so that $\ket{g_A}=\ket{g_B}$,
$\ket{e_A}=\ket{e_B}$,  $\Omega_A=\Omega_B$, and likewise for the switching function
$\chi_A(\tau) =\chi_B(\tau) =\exp(-\tau^2/2\sigma^2)$, where $\sigma$ is the characteristic timescale of the Gaussian.

The evolution of the whole system in the Schr{\"o}dinger picture, including the free evolution of the black hole governed by $\hat{H}_{0,S}$, is given by 
\begin{equation}
    \ket{\psi(t_{f})}=e^{-i\hat{H}_{0,S}t_f}\hat{U}e^{i\hat{H}_{0,S}t_i}\ket{\psi(t_{i})} ,
    \label{eq: final state}
\end{equation}
where the time-evolution operator $\hat{U}$ can be written in the usual manner, 
\begin{align}
    \hat{U} &= \mathcal{T} \exp \left( - i \int_{t_i}^{t_f} \D \tau \: \hat{H}_\mathrm{int}(\tau) \right) ,
\end{align}
where $\mathcal{T}$ denotes time-ordering. For simplicity, it is convenient for us to absorb the phases arising from the free-evolution into the definition of the states, a technique commonly performed in atomic physics known as a ``rotating frame transformation'' \cite{PhysRevA.29.1188,winelandPhysRevA.50.67}. That is, the evolution of the black hole in a superposition of energy eigenstates under the free Hamiltonian $\hat{H}_\mathrm{0,BH} = E_A | M_A, J_A \rangle\langle M_A , J_A | + E_B | M_B , J_B \rangle\langle M_B , J_B |$, will in general introduce a time-dependent oscillation in the final transition probability of the detector, and was previously explored in \cite{Foo_2022}. Here, we will focus on the indirect influence of the black hole on the detector through its interaction with the field.

Given our assumption of a weak interaction $(\lambda \ll 1)$ it is convenient for us to compute the evolution of the state perturbatively. To second-order in $\lambda$, the operator $\hat{U}$ takes the form 
\begin{equation}
    \hat{U}=I+\hat{U}^{(1)}+\hat{U}^{(2)}+\mathcal{O}(\lambda^3)
\end{equation}
where $\hat{U}^{(1)},\hat{U}^{(2)}$ are respectively the first and second order term of the evolution operator
\begin{align}
    \hat{U}^{(1)}&=-i\lambda\int_{t_i}^{t_f}\D \tau \: \hat{H}_{int}(\tau)\\
    \hat{U}^{(2)}&=-\lambda^2\int_{t_i}^{t_f}\D \tau\int_{t_i}^\tau \D \tau' \: \hat{H}_{int}(\tau) \hat{H} _{int}(\tau'),
\end{align}
where $t_i,t_f$ are the initial and final times of the interaction. Following the interaction, we assume that a conditional measurement on the final state of the black hole can be made, which for simplicity we consider to be the $| + \rangle = ( | M_A , J_A \rangle + | M_B , J_B \rangle )/\sqrt{2}$ basis. Such a measurement corresponds to a recombination of paths in a Mach-Zehnder interferometer, and could in-principle be performed by entangling an ancilla with the black hole and later projecting the ancilla onto the corresponding superposition state. The conditional state of the detector and field is given by, 
\begin{align}
    \langle + | \hat{U} | \psi(t_i) \rangle &= \frac{1}{2} ( \hat{U}_A + \hat{U}_B ) | \psi(t_i) \rangle 
    \vt 
\end{align}
where $\hat{U}_A$ and $\hat{U}_B$ are the separate evolutions governed by the interaction Hamiltonians corresponding to $D = A, B$ respectively in Eq.\ (\ref{eq9}). Upon tracing out the field degrees of freedom, the resulting conditional transition probability of the detector $P_T \equiv P(g \to e)$ is given by
\begin{align}
    P_T &= \frac{1}{4} ( P_A + P_B + 2L_{AB} ) + \mathcal{O}(\lambda^4)\label{eq:transprobability}
\end{align}
where 
\begin{align}
     P_D &= \lambda^2\int_{-\infty}^\infty \D \tau \D \tau' e^{- \frac{\tau^2}{2\sigma^2}} e^{- \frac{\tau'^2}{2\sigma^2}} e^{-i\Omega(\tau - \tau' ) } W_\mathrm{BTZ}^{(D)}(x, x') \label{PDgeneral}
\\
    L_{AB} &= \lambda^2\int_{-\infty}^\infty \D \tau \D \tau' e^{- \frac{\tau^2}{2\sigma^2}} e^{- \frac{\tau'^2}{2\sigma^2}} e^{-i\Omega(\tau - \tau' ) } W_\mathrm{BTZ}^{(AB)}(x, x') \label{LABgeneral}
\end{align}
Here, $P_D$ corresponds to the transition probability of the detector if it were situated in the classical spacetime corresponding to mass $M_D$ and angular momentum $J_D$. Meanwhile, $L_{AB}$ is a cross-correlation term that features the Wightman function evaluated with respect to fields quantized on both spacetimes, $M_A,J_A$ and $M_B,J_B$. In the context of entanglement harvesting \cite{Robbins_2022}, a functionally equivalent term is interpreted as a measure of correlations between two detectors along different worldlines in the same classical spacetime. Here, it arises due to the interference between the two spacetimes.

\subsection{Trajectory of the Detector}

As mentioned earlier, for ease of computational analysis we consider the detector to be co-rotating with the black hole 
\cite{PhysRevD.86.064031,Robbins_2022}.
Specifically, we consider the case where the trajectory of the detector is entangled with the black hole--it moves along a superposition of trajectories \cite{Foo_2020,fooPhysRevResearch.3.043056} in a way that allows us to naturally define the time-evolution parameter with respect to a clock at large radial distance from either of the superposed horizons (as shown below, Sec.\ \ref{secE}). In the case of a single black hole, this corresponds to \cite{PhysRevD.86.064031}
\begin{align}
    t &= \frac{l  r_+ \tau}{\sqrt{r^2 - r_+^2 }\sqrt{r_+^2 - r_-^2}} \\
    \phi &= \frac{r_- \tau}{\sqrt{r^2 - r_+^2}\sqrt{r_+^2 - r_-^2}}\,,
\end{align}
where $(t,\phi)$ corresponds to the coordinates of the detector. For a mass-superposed black hole, the 
respective coordinates in
each of the co-rotating frames (corresponding to each mass) are
\begin{align}\label{t-and-phi-coord}
    t_A &= \frac{lr_{+A}\tau}{\sqrt{r^2  - r_{+A}^2}\sqrt{r_{+A}^2 - r_{-A}^2}} \\
    t_B &= \frac{lr_{+B} \tau'}{\sqrt{r^2 - r_{+B}^2}\sqrt{r_{+B}^2 - r_{-B}^2}} \\
    \phi_A &= \frac{r_{-A}\tau}{\sqrt{r^2 - r_{+A}^2}\sqrt{r_{+A}^2 - r_{-A}^2}} \\
    \phi_B &= \frac{r_{-B} \tau'}{\sqrt{r^2 - r_{+B}^2}\sqrt{r_{+B}^2 - r_{-B}^2}} \label{t-and-phi-coord2} 
\end{align}
We fix the detector at radial coordinate $r=R$  yielding 
\begin{align} 
    \gamma_D=\frac{\sqrt{R^2-r_{+D}^2}\sqrt{{r_{+D}}^2-{r_{-D}}^2}}{lr_{+D}}
    \label{gammaD}
\end{align}
for the redshift factor in each of the frames $D=A,B$. In summary, the detector trajectory is defined via fixed radial coordinate $R$ and proper time $\tau$ independent of the state of the black hole, thus the coordinate time as well as the angular coordinate depend on the state of the black hole as shown in Eqs~\eqref{t-and-phi-coord}--\eqref{t-and-phi-coord2}, and the trajectory of the detector is thus  correlated with the state of the black hole.

\subsection{Transition Probability of the Detector}\label{secE}

Explicitly inserting the Wightman functions, Eq.\ (\ref{eq4}), into Eqs\ (\ref{PDgeneral}) and (\ref{LABgeneral}), we find, as shown in Appendix \ref{Appendix:Appendix}, that
\begin{widetext} 
\begin{align}
    \frac{P_D^{(\zeta)}}{\lambda^2\sigma} &= \frac{1}{\mathcal{N}} \sum_{n,m} \mathrm{Re} \int_0^\infty\D z \: \frac{e^{- \bar{z}_{nm}^2/4\sigma^2Y_D^2l^2}e^{- i \Omega \bar{z}_{nm}/Y_Dl}}{2\sqrt{2\pi}}  \left[ \frac{1}{ A_{nm}^{(+ D)}(z)} - \frac{\zeta}{ A_{nm}^{(- D)}(z)} \right] \label{PDzeta-T} 
\\
    \frac{L_{AB}^{(\zeta)}}{\lambda^2\sigma} &= \frac{e^{-\frac{\Upsilon_-^2\sigma^2\Omega^2}{2\Upsilon}}}{\sqrt{\boldsymbol{\alpha}} \sqrt{\Upsilon}} \frac{1}{\mathcal{N}}\sum_{n,m} \mathrm{Re} \int_0^\infty \D z \: \frac{ e^{- \frac{(z+\beta_{nm})^2}{2\Upsilon\sigma^2}} e^{-\frac{i\Omega \Upsilon_+}{\Upsilon} \left( z + \beta_{nm} \right)} }{2\sqrt{\pi} l} \left[ \frac{1}{ A_{nm}^{(+ X)}(z)} - \frac{\zeta}{ A_{nm}^{(- X)}(z)} \right]\label{LABzeta-T}
\end{align}
where
\begin{align}
\bar{z}_{nm} &= z - 2\pi r_{-D}(m-n)/l
\qquad 
\Upsilon = Y_A^2 + Y_B^2
\qquad
\Upsilon_\pm = Y_A \pm Y_B \qquad
    \boldsymbol{\alpha} = \sqrt{( \alpha_A - 1)( \alpha_B - 1 )}
    \vphantom{\frac{1}{2}} 
\\
    A_{nm}^{(\pm D)}(z) &= \sqrt{\cosh\alpha_{nm}^{(\pm D)} - \cosh(z)} 
    \qquad
    A_{nm}^{(\pm X)}(z)= \sqrt{\cosh \alpha_{nm}^{(\pm X)} - \cosh(z) }
    \vphantom{\frac{1}{2}} \label{eqs:Anm}
\\
    \alpha_D &=\frac{R^2-r_{-D}^2}{r_{+D}^2-r_{-D}^2}\qquad
    \beta_{nm} = \frac{2\pi}{l}(nr_{-A} - m r_{-B})\qquad
    Y_D = \frac{\sqrt{r_{+D}^2 - r_{-D}^2}}{l\sqrt{R^2 - r_{+D}^2}}
\\
    \cosh\alpha_{nm}^{(\pm D)}&=\frac{\alpha_D}{\alpha_D-1}\cosh\left[ \frac{2\pi r_{+D}(m-n)}{l} \right] \mp \frac{1}{\alpha_D-1}\\
    \cosh\alpha_{nm}^{(\pm X)} &= \frac{\sqrt{\alpha_A \alpha_B}}{\boldsymbol{\alpha}} \cosh \left[ \frac{2\pi(nr_{+A} - m r_{+B} )}{l} \right] \mp \frac{1}{\boldsymbol{\alpha}}\label{coshX}
\end{align}
\end{widetext} 

For convenience we will henceforth fix the variables $l=5\sigma, M_A=1$. By setting up the parameters with units of $\sigma,\lambda$, the scaling of $\sigma\lambda$ only changes the transition probability by a numerical factor which for our purposes results in the same conclusion. We computationally show that the sums in  Eqs~\eqref{PDzeta-T},~\eqref{LABzeta-T} converge with respect to truncated maximum values of $n,m$ in Appendix \ref{Appendix:nmax}.

We remark here that to obtain Eq.~\eqref{LABzeta-T}, we have utilized a variable transformation to a common time coordinate, $z = T - T'$, where $T = Y_A \tau,  T' = Y_B \tau'$ (by symmetry, a similar transformation is used for the $L_{BA} = L_{AB}$ term). The physical interpretation of such a transformation is the use of a clock at large radial distances from either of the superposed horizons, 
such that the fractional difference between the proper time in the respective amplitudes of the spacetime is 
\begin{align} 
    \frac{(M_B-M_A)l^2}{2r}+\frac{(J_A^2-J_B^2)l^2}{8r^3} .
\end{align}
This is approximately zero at for sufficiently large $r$; likewise, from
Eq.~\eqref{metric} the metric is approximately static.

\section{Superposed Black Holes of Equal Angular Momentum}\label{sec:Equal angular momentum}

\subsection{General Boundary Conditions}

We first generalize the results of Ref.\ \cite{Foo_2022} $(J = 0)$ to include both Neumann $(\zeta = -1)$ and Dirichlet $(\zeta = 1)$ boundary conditions (of the field at spatial infinity), in addition to the transparent boundary condition $(\zeta = 0)$, which interpolates between Neumann and Dirichlet. In Fig.~\ref{nonrotating}, we  plot the transition probability with respect to the ratio between the outer radii of the two black hole states 
\begin{align} 
    \mathcal{R}:=r_{+B}/r_{+A}
\nonumber 
\end{align}
with the vertical gridlines (for e.g.\ at $1/3,2/5, 1/2$ and so on), marking some of these rational values of $\mathcal{R}$ along the domain where $P_T$ resonates. We find that the values of $\mathcal{R}$ corresponding to the resonance locations  in \autoref{nonrotating} are preserved for all boundary conditions, with the overall magnitude of $P_T$ decreasing as $\zeta$ increases from negative to positive values. We see that the consistency of these  resonances  with Bekenstein's conjecture is robust across field boundary conditions.

The transition probability reaches a global maxima at $\mathcal{R}=1$, corresponding to the case where the black hole masses are equal.  This in turn implies that the black holes are identical. In this case there is no superposition, since the spacetimes are indistinguishable.
 
The resonant nature of the plots 
can be attributed to the $\cosh{\alpha_{nm}^{(\pm X)}}$ term in the denominator of $L_{AB}^{(\zeta)}$, see Eqs~\eqref{LABzeta-T} and \eqref{eqs:Anm}, 
which is minimized when $\mathcal{R}$
is rational, as seen in Eq.~\eqref{coshX}. This argument applies to all boundary conditions, ensuring that  the locations of the resonances remain the same for all values of $\zeta$.

\begin{figure}[H]
    \centering
    \includegraphics[width=1\linewidth]{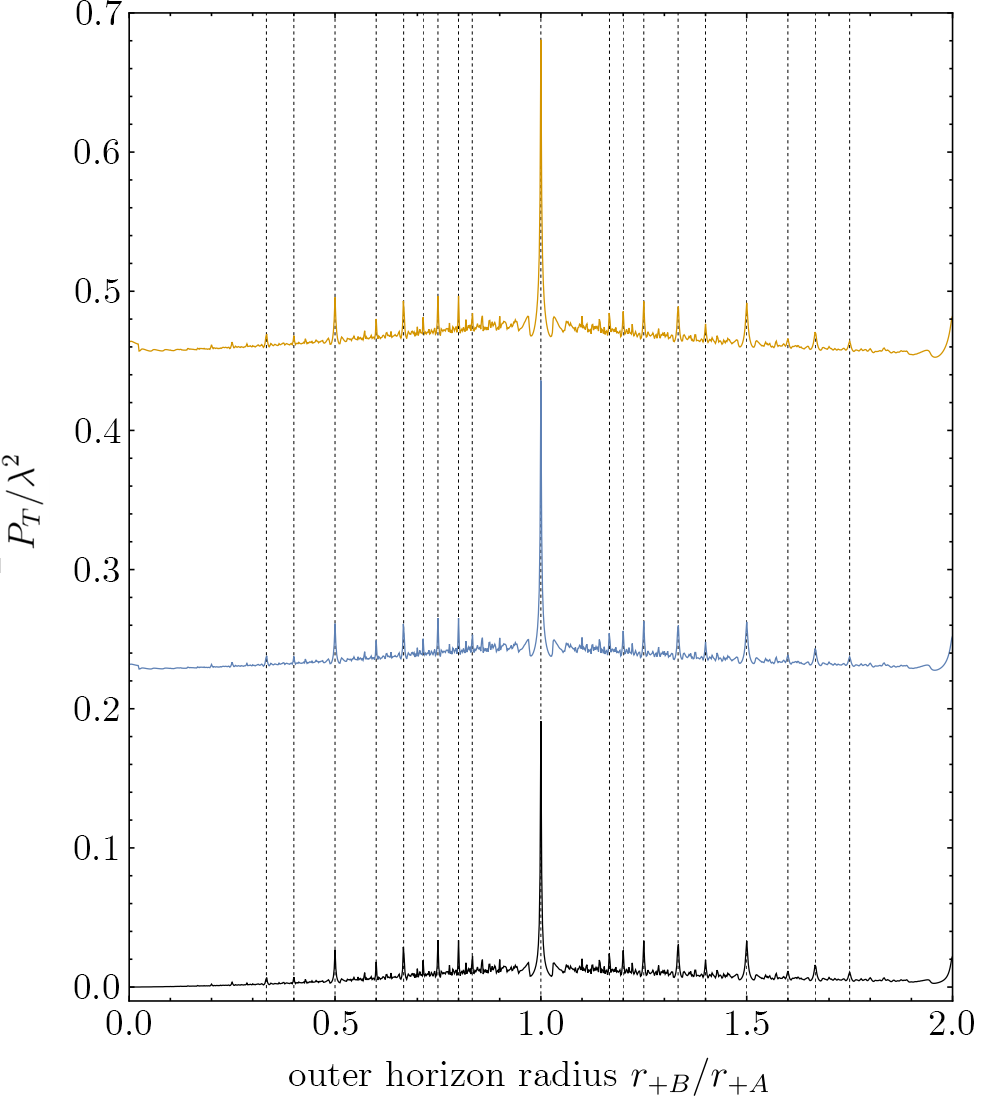}
    \caption{Plot of the transition probability as a function of the ratio of the outer radii of two superposed nonrotating $(J/l=0)$ states of the black hole $A$ and $B$ with parameter $l=5\sigma, R/l=3, M_A=1, \sigma\Omega=0.02$. The different colors correspond to $\zeta = -1$ (yellow, Neumann), $\zeta = 0$ (blue, transparent), and $\zeta = 1$ (black, Dirichlet). Vertical gridlines were added for distinctive peaks at rational values (e.g $1/3, 2/5, 1/2$). }
    \label{nonrotating}
\end{figure}

\subsection{Nonzero Angular Momenta}\label{secIIIB}

\begin{figure*}
    \centering
    \includegraphics[width=\linewidth]{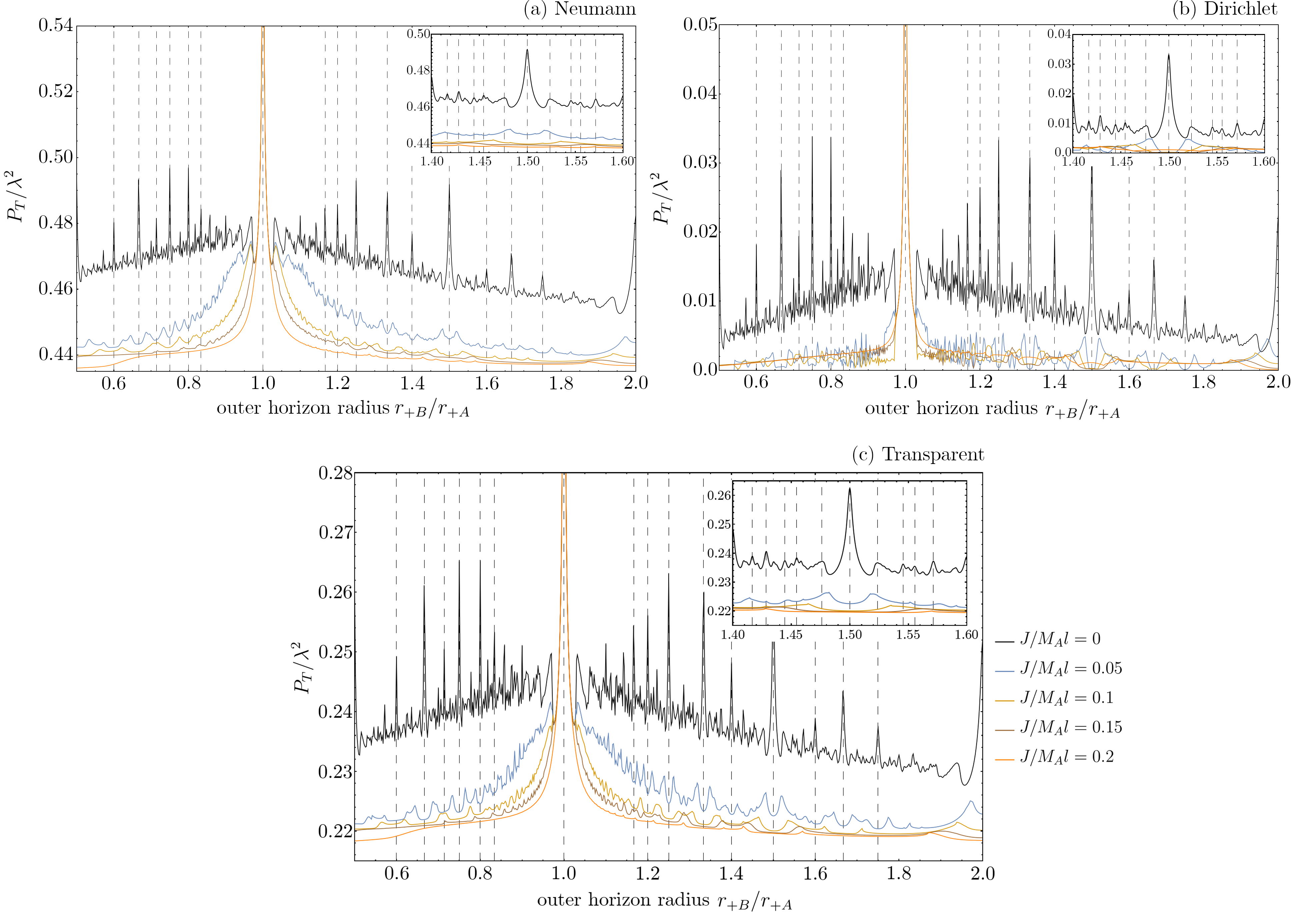}
    \caption{Plot of the transition probability as a function of the outer horizon ratio of the two states of the rotating black hole $A$ and $B$  with parameters $l=5\sigma, R/l=3, M_A=1, \sigma\Omega=0.02$ and equal parametric angular momentum $J$.  The dashed vertical lines are at rational values of $\mathcal{R}$ (e.g at $1/3, 2/5, 1/2$). As the angular momentum $J$ increases, the resonant peaks split and the  curves diminish. As $J$ further increases, the curves become smoother, with the magnitudes of the  resonances rapidly approaching zero (except for the central peak). The insets show a more detailed graph of the transition probability for the domain $1.4\leq\mathcal{R}\leq1.6$.}
\label{fig:variedJ}
\end{figure*}

We now consider nonzero values 
for the angular momenta of the superposed black holes, taking $J_A=J_B=J$.  
In \autoref{fig:variedJ}, we plot the transition probability $P_T$ 
as a function of  $\mathcal{R}$ for various values of $J$ and all boundary conditions. We observe that as $J$ increases, the resonant peaks split and rapidly diminish in magnitude,
eventually disappearing for sufficiently large $J$ (for example  $J/M_Al=0.2$ in \autoref{fig:variedJ}) with the exception of the central peak. Furthermore,
$P_T$ becomes an increasingly smooth function of $J$. 

We attribute this behaviour to the additional presence of the  
  $\beta_{nm}$ term in Eq.~\eqref{LABzeta-T}. 
  This term exponentially suppresses the quantum interference term. Indeed this term and the term in Eq.~\eqref{coshX} both suppress $L_{AB}$ unless they simultaneously vanish. This will occur for 
\begin{align}
    n r_{\pm A} - m r_{\pm B} = 0 
    \vphantom{\frac{1}{\sqrt{2}}}
    \nonumber 
\end{align}  
respectively. Resonances that occur for $nr_{+A}-mr_{+B}=0$ (as in the static case) will be suppressed if $(l/2\pi) \beta_{nm} = nr_{-A}-mr_{-B} \neq 0$. It is therefore of interest to investigate situations in which $\beta_{nm} = 0$ at resonance.  This we do in the following section.

Finally, we note that by fixing the value of $M_A$ and $J$, and imposing that $r_{-D}\leq r_{+D}\leq R$, we obtain (as shown in Appendix \ref{Appendix:DomainR}) the following domain for $\mathcal{R}$
\begin{equation}
   \sqrt{\frac{2}{Jl}}r_{-A}\leq \mathcal{R}\leq \frac{2R}{Jl}r_{-A}\label{gammaJrange}.
\end{equation}

\begin{figure*}
    \centering
    \includegraphics[width=0.9\linewidth]{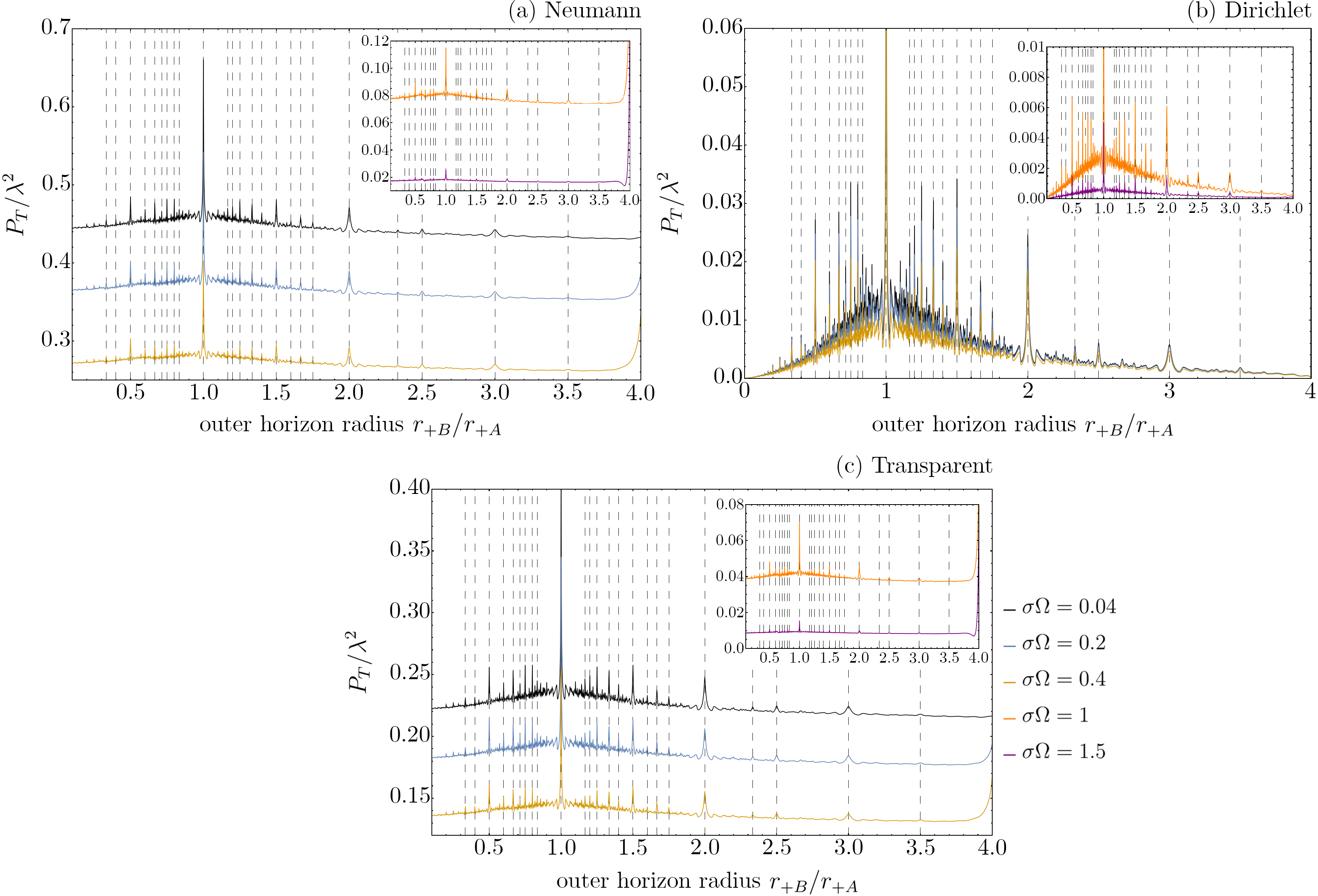}
    \caption{Plot of transition probability as a function of the outer horizon radius ratio $\mathcal{R}$ at $\varrho_A=\varrho_B=3$ for all boundary \textcolor{white}{aaaaaaa} conditions  and various energy gaps. Here $l=5\sigma, R/l=3, M_A=1$ and varied energy gap $\sigma\Omega=0.04,0.2,0.4$. The inset \textcolor{white}{aaaa} represents energy gap $\sigma\Omega=1,1.5$. We see, for all boundary conditions, that an increase in the energy gap  decreases the \textcolor{white}{aaaaaa} magnitude of $P_T$ whilst preserving its resonance structure. \textcolor{white}{aaaaaaaaaaaaaaaaaaaaaaaaaaaaaaaaaaaaaaaaaaaaaaaaaaaaaaaaaaaaa}}
    \label{fig:Omegavaryplot}
\end{figure*}


\section{Resonances of Superposed Black holes}\label{sec:fixed ratio}

In the previous section, we showed that the so-called ``Bekenstein resonances'' in the detector transition probability become exponentially suppressed when the black hole possesses a nonzero and equal angular momenta in each amplitude of the superposition. We are interested in scenarios in which these resonances persist, possibly for nontrivial combinations of the angular momentum in each amplitude. 

\noindent To perform our analysis, let us first define the ratios, 
\begin{align}
    \varrho_A &= \frac{r_{+A}}{r_{-A}} \geq 1 
    \nonumber \\
    \varrho_B &= \frac{r_{+B}}{r_{-B}} \geq 1 
    \nonumber 
\end{align}
of the outer and inner horizon radii. Since the  detector is  outside the black hole, the following conditions 
\begin{align}\label{rangeR}
    R &\geq \varrho_Ar_{-A} 
    \nonumber 
    \vphantom{\frac{1}{\sqrt{2}}} 
    \\
    \mathcal{R} &=\frac{r_{+B}}{r_{+A}}\leq\frac{R}{\varrho_A r_{-A}}
    \nonumber 
\end{align}
must also be satisfied. In the following, we are primarily interested in the behaviour of the detector as a function of $\mathcal{R}$, the ratio of the outer horizons of the superposed spacetime. 

We consider first equal values of 
$\varrho_A$ and $\varrho_B$. The interdependence of $\varrho_A, \varrho_B, \mathcal{R}$ on the angular momentum of each amplitude $(J_A, J_B)$ means that in general these will not be fixed as $\mathcal{R}$ is varied. We see 
from \autoref{fig:Omegavaryplot}  
that the resonance structure in the static case is present for all boundary conditions and a broad range of energy gaps $\sigma\Omega$.  As in the static case, a comparison of the various graphs at any given $\sigma\Omega$ shows that the response decreases as $\zeta$ increases from $-1$ to $1$. In addition, the magnitude of the response monotonically decreases as $\sigma\Omega$ increases, with the structure of the resonances preserved. The relative change in magnitude likewise decreases, being least pronounced for Dirichlet boundary conditions.

\autoref{fig:RDvaryplot} shows how the transition probability changes as we increase the radial coordinate $R$} of the detector. For small values of the outer horizon ratio, the general trend of the response is to  initially decrease (except in the Dirichlet case), then increase until it reaches a maximum at $\mathcal{R}=1$, after which  it  
monotonically decreases again, with resonances punctuating the curve at rational values of $\mathcal{R}$.
At large values of the detector's radial location $R$ the curves  converge.  This convergence can be interpreted in terms of a diminishing effect that the (superposed) black hole has on the detector (recall the spacetime is asymptotically AdS), apart from the persistent resonances due to   the space-time identification.
 
\begin{figure*}
    \centering
    \includegraphics[width=\linewidth]{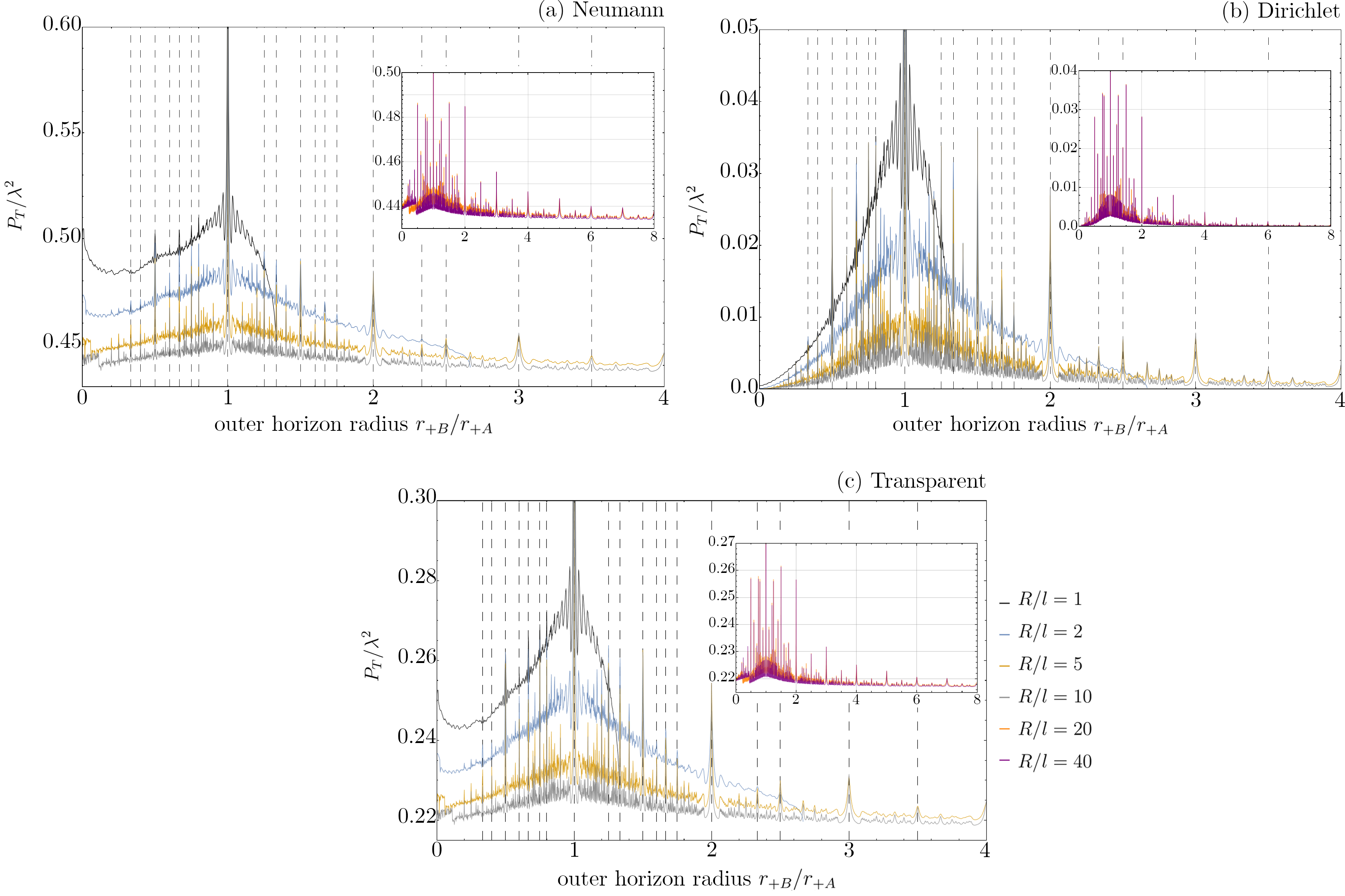}
    \caption{Multiline plot of transition probability against $\mathcal{R}=r_{+B}/r_{+A}$ at different values of $\varrho_A=\varrho_B=3$, at $R/l=1,2,5,10,20$.  For convenience and comparability, the values of $R/l$ are chosen so that some of them coincide with previously generated data. As the detector radius increases, the domain of the inner horizon radii increases and the mean values of the curves converge (modulo peaks). The lines appear to suddenly drop near the end of its domain (characterized by Eq.~\eqref{rangeR}), where it intersect with the lines corresponding to the other parameters.  The inset represents the plot of transition probability at $R/l=20, 40$. The fixed parameters are such that previously generated data can be reused (e.g \autoref{fig:Omegavaryplot} also utilize the $R/l=3$ condition). The sudden drop at low values of $\mathcal{R}=r_{+B}/r_{+A}$ is a result of computational inaccuracy and should be disregarded. Vertical gridlines have been placed at rational values  of $\mathcal{R}$.}
\label{fig:RDvaryplot}
\end{figure*}

\begin{figure*}
    \centering
    \includegraphics[width=\linewidth]{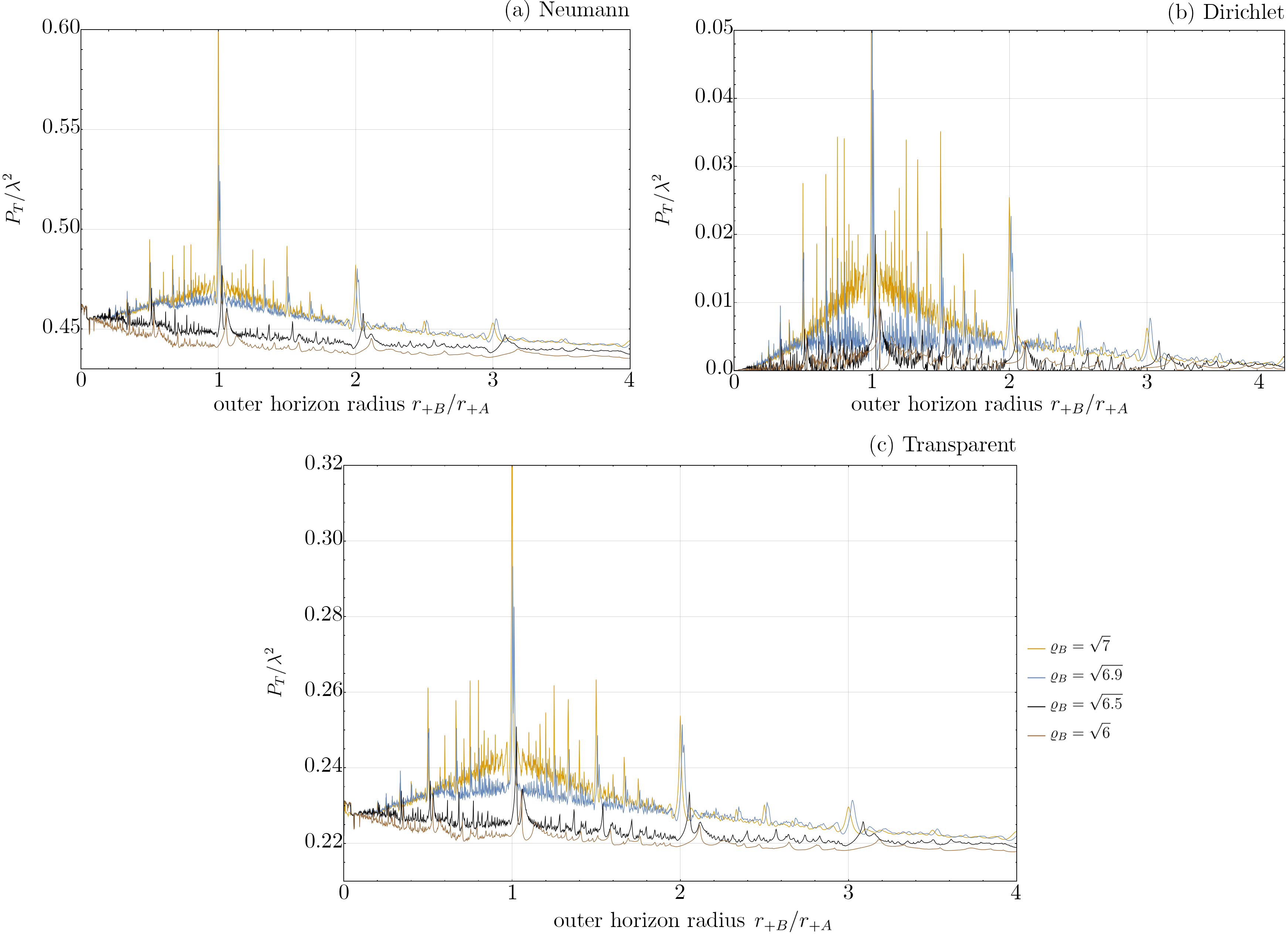}
    \caption{Plot of transition probability as a function of $\mathcal{R}=r_{+B}/r_{+A}$ at $\varrho_A=\sqrt{7}$, $l=5\sigma, R/l=3, M_A=1,r_{-A}/l=0.25 ,$ $\sigma\Omega=0.02$ and parametric $\varrho_B=\sqrt{7},\sqrt{6.9},\sqrt{6.5},\sqrt{6}$ at different boundary conditions $\zeta=-1,1,0$. We have chosen $\varrho_A$ and $\varrho_B$ to have arbitrary irrational values to illustrate that this general behaviour does not depend on $\varrho$ being a rational number for either black hole state. As the two ratios $\varrho_A$ and $\varrho_B$ deviate from each other, the local maxima become split in two and are \textcolor{white}{aa} considerably dampened.\textcolor{white}{aaaaaaaaaaaaaaaaaaaaaaaaaaaaaaaaaaaaaaaaaaaaaaaaaaaaaaaaaaaaaaaaaaaaaaaaaaaaaaaaaaaaaaaaaaa}}
    \label{fig:RTBvaryplot}
\end{figure*}


Turning next to differing inner horizon ratios, in \autoref{fig:RTBvaryplot} we explore how the constructive interference changes as the ratio $\varrho_B$ varies for fixed  $\varrho_A$.   We see that as  $\varrho_B$ increasingly deviates from $\varrho_A$, the overall response curve diminishes  while the peaks split. This reduction happens rapidly, with the maxima of the probability occurring at larger values of $\mathcal{R}$ as $\varrho_B$ decreases. Though we have chosen an arbitrary irrational value for $\varrho_A$, this behaviour does not depend on that choice.

In a similar way to the result shown in Sec.\ \ref{secIIIB}, this behaviour can be attributed to the increasing value of 
$\beta_{nm}$. 
When $\varrho_A=\varrho_B$, the denominator term, Eq.\ \eqref{coshX}, in Eq.\ \eqref{LABzeta-T} is minimized (i.e $nr_{+A}-mr_{+B}=0$) while the Gaussian factor
\begin{align}
    \exp \left[- \frac{1}{2(Y_A^2 + Y_B^2) \sigma^2} \left( z + \beta_{nm} \right)^2\right]
\end{align}
in Eq.\ \eqref{LABzeta-T} is maximized (i.e $nr_{-A}-mr_{-B}=0$). Hence, constructive interference occurs at  rational values of $\mathcal{R}=r_{+B}/r_{+A}$. As the ratio $\varrho_A$ and $\varrho_B$ deviate from unity, $\beta_{nm}$ increases and
the constructive interference becomes smaller, resulting in dampening and peak splitting. Finally, we note that the graphical discontinuity at very low values of $\mathcal{R}$ in \autoref{fig:RDvaryplot} are results of numerical convergence errors and can be disregarded.

\section{Conclusion}\label{sec:conclusion}

We have shown that, under certain conditions, 
rotating black holes in superposition 
cause an Unruh-deWitt detector to exhibit a resonant response rate at rational values of the ratio $\mathcal{R}$ of the outer horizon radii.  The effect is clearly present if the ratio of outer to inner horizon radius is equal for the two states of the black hole. This resonant effect is very robust: it is present for a broad range of detector energy gaps, a broad range of detector orbits, and for all boundary conditions of the conformally coupled scalar field.

When the black hole is not rotating we recover previous results for the static case \cite{Foo_2022}. As for that case,  we conclude that  the detector is in a sense ``sensitive'' to a version of the discretization of a black hole horizon, consistent with Bekenstein's conjecture.

We demonstrated that the presence of the ``Bekenstein resonances'' remained when the ratio between the inner and outer horizon radii in the respective amplitudes was equal. However when difference between these ratios increased, the resonances ``split'' in two, and eventually vanished altogether. Another interesting result was that though increases in the detector energy reduced the detector response, this effect was shown to be unexpectedly much subtler in the Dirichlet case, in contrast with the Neumann and transparent boundary conditions. We emphasize that the presence or absence of these resonances arises due to likely nontrivial mode structure of the fields quantized on the superposed amplitudes of the spacetime. Here, this interference is encoded within the two-point correlation (Wightman) functions that are integrated over the interaction period between the detector, black hole, and field. To obtain a more detailed understanding of this interference behaviour, one should analyze the mode structure of the field (superpositions of fields) from first principles, in particular its dependence on the value of the horizon radii (ratio of the horizon radii). We leave this analysis for a future work.


As the detector is placed further away from the black hole, its overall response with respect to $\mathcal{R}$  decreases, as expected, until it converges  to some mean value (modulo peaks).
This  convergence can be interpreted as the diminishing effect of the superposition as the detector is placed further away, until almost all its response becomes solely a result of the spacetime identification. 

More broadly, we have provided further supporting evidence that rotating black holes can be discretely quantized in terms of their outer horizon radius. This provides an important link between the quantum-gravitational effects that are derivable using the tools of relativistic quantum information and curved spacetime quantum field theory, with those obtained from a phenomenological quantum gravity perspective, namely the Bohr-Sommerfeld quantization scheme first applied by Bekenstein to black holes.

\section{Acknowledgements}
This work was supported in part by the Natural Sciences and Engineering Research Council of Canada. J.F.\ is supported by funding provided by the U.S. Department of Energy, Office of Science, ASCR under Award Number DE-SC0023291.  M.Z.\ is supported by Knut and Alice Wallenberg foundation through a Wallenberg Academy Fellowship No.\ 2021.0119

\appendix
\begin{widetext}

\section{Calculating $P_D^{(\zeta)},L_{AB}^{(\zeta)}$}\label{Appendix:Appendix}
\noindent The geodesic distances $\sigma(\cdot,\cdot)$ between the spacetime events $\Gamma_D^n x , \Gamma_D^m x'$ $(D = A, B)$ and $\Gamma_A^n x, \Gamma_B^m x'$ in the BTZ spacetime are respectively
\begin{align}
    & \sigma(\Gamma_D^n x, \Gamma_D^m x') 
    \nonumber \vt \\
    & \:\:\: = \left( \alpha_D -1 \right) \left[ \frac{\alpha_D}{\alpha_D-1} \cosh\left[ \frac{2\pi r_+(m-n)}{l} \right] - \frac{1}{\alpha_D-1} - \cosh \left[ \frac{\big( r_+^2 - r_-^2 ) (\tau -\tau')}{\Lambda_D} + \frac{2\pi r_{-D}(m-n)}{l} \right]  \right]\label{sigmaD}
\\
    & \sigma(\Gamma_A^n x, \Gamma_B^m x') \nonumber 
    \vt \\
    & \:\:\: = \sqrt{(\alpha_A -1 )( \alpha_B - 1) } \bigg[ \frac{\sqrt{\alpha_A \alpha_B}}{\sqrt{(\alpha_A - 1) ( \alpha_B - 1 ) }} \cosh \left[ \frac{2\pi (nr_{+A} - m r_{+B} ) }{l} \right] - \frac{1}{\sqrt{(\alpha_A - 1 )( \alpha_B - 1) }}  \label{sigmaAB}\\
    & - \cosh \left[ Y_A \tau - Y_B \tau' - \frac{2\pi}{l}( n r_{-A} - m r_{-B} ) \right] \bigg]\notag
\end{align}
The Wightman function in AdS$_3$ is given by
\begin{equation}
    W^{(\zeta)}(x,x')=\frac{1}{4\pi l\sqrt{2}}\left(\frac{1}{\sqrt{\sigma(x,x')}}-\frac{\zeta}{\sqrt{\sigma(x,x')+2}}\right)\label{zeta wightman}
\end{equation}
To obtain the Wightman functions in the BTZ spacetime, one utilizes the method of images, giving
\begin{align}
    W^{(D)}_\mrm{BTZ}(x,x')=\frac{1}{4\pi l\sqrt{2}}\frac{1}{\mathcal{N}}\sum_{n,m}\left(\frac{1}{\sqrt{\sigma(\Gamma_D^nx,\Gamma_D^nx')}}-\frac{\zeta}{\sqrt{\sigma(\Gamma_D^nx,\Gamma_D^nx')+2}}\right)\label{wightmanD}
\\
    W^{(AB)}_\mrm{BTZ}(x,x')=\frac{1}{4\pi l\sqrt{2}}\frac{1}{\mathcal{N}}\sum_{n,m}\left(\frac{1}{\sqrt{\sigma(\Gamma_A^nx,\Gamma_B^nx')}}-\frac{\zeta}{\sqrt{\sigma(\Gamma_A^nx,\Gamma_B^nx')+2}}\right)\label{wightmanAB}
\end{align}
Here $\mathcal{N}:=\sum_n\eta^{2n}$ is the overall normalization which preserves the canonical commutation relation,  $[ \phi(x) , \phi(x')] = \delta(x - x') + \mathrm{image \: terms}$. Inserting Eqs~\eqref{sigmaD} and \eqref{sigmaAB} to Eqs~\eqref{wightmanD} and \eqref{wightmanAB} we obtain
\begin{align}
    W^{(D)}_\mrm{BTZ}(x,x') &=\frac{1}{4\pi l\sqrt{2}}\frac{1}{\mathcal{N}}\frac{1}{\sqrt{\boldsymbol{\alpha}}}\sum_{n,m}\Biggr[ \frac{1}{\sqrt{\cosh\alpha_{nm}^{(+ D)} - \cosh \left[ Y_D(\tau-\tau') + \frac{2\pi r_{-D}(m-n)}{l} \right]}}\\& - \frac{\zeta}{\sqrt{\cosh\alpha_{nm}^{(- D)} - \cosh \left[ Y_D(\tau-\tau') + \frac{2\pi r_{-D}(m-n)}{l} \right]}} \Biggr]\label{wightmanD2}
\\
    W^{(AB)}_\mrm{BTZ}(x,x') &=\frac{Y_D}{4\pi l\sqrt{2}}\frac{1}{\mathcal{N}}\sum_{n,m}\Biggr[ \frac{1}{\sqrt{\cosh\alpha_{nm}^{(+ X)}  - \cosh \Big[ Y_A \tau - Y_B \tau' - \beta_{nm}\Big]}}\label{wightmanAB2}\\& - \frac{\zeta}{\sqrt{\cosh\alpha_{nm}^{(- X)} -  \cosh \Big[ Y_A \tau - Y_B \tau' - \beta_{nm} \Big]}} \Biggr]\notag
\end{align}
where
\begin{align}
    \cosh\alpha_{nm}^{(\pm D)}&=\frac{\alpha_D}{\alpha_D-1}\cosh\left[ \frac{2\pi r_{+D}(m-n)}{l} \right] \mp \frac{1}{\alpha_D-1}\\
    \cosh\alpha_{nm}^{(\pm X)} &= \frac{\sqrt{\alpha_A \alpha_B}}{\boldsymbol{\alpha}} \cosh \left[ \frac{2\pi(nr_{+A} - m r_{+B} )}{l} \right] \mp \frac{1}{\boldsymbol{\alpha}}
\end{align}
and 
\begin{equation}
    \boldsymbol{\alpha} = \sqrt{( \alpha_A - 1)( \alpha_B - 1 )}
    \vphantom{\frac{1}{2}} \qquad 
    \alpha_D =\frac{R^2-r_{-D}^2}{r_{+D}^2-r_{-D}^2}\qquad
    \beta_{nm} = \frac{2\pi}{l}(nr_{-A} - m r_{-B})
\end{equation} 
\begin{equation}
     Y_D = \frac{\sqrt{r_{+D}^2 - r_{-D}^2}}{l\sqrt{R^2 - r_{+D}^2}}=\frac{1}{\sqrt{\alpha_D-1}}
\end{equation}
\begin{align}
    \boldsymbol{\alpha}^{(\pm D)} = \sqrt{\cosh\alpha_{nm}^{(\pm D)} - \cosh(z)} 
    \qquad
    \boldsymbol{\alpha}^{(\pm X)} = \sqrt{\cosh \alpha_{nm}^{(\pm X)} - \cosh(z) }
    \vphantom{\frac{1}{2}} 
\end{align}
Explicitly inserting the Wightman function in Eq.~\eqref{wightmanD2} to Eq.~\eqref{PDgeneral} results in a double integral with respect to $\tau,\tau'$. Because $W^{(D)}_\mrm{BTZ}(x,x')$ is only dependant on $\tau-\tau'$, we can substitute the integration variables for $z=Y_D(\tau-\tau'), u=Y_D(\tau+\tau')$ and integrate with respect to $u$ (this is easily calculated as we are simply integrating a Gaussian). Hence we obtain
\begin{align}
P_D^{(\zeta)}  &=\frac{\sigma}{4l\sqrt{2\pi}} \frac{1}{\mathcal{N}} \sum_{n,m} \int_{-\infty}^\infty \D z\: e^{- z^2/4\sigma^2Y_D^2} e^{- i\Omega z/Y_D} \Biggr[\frac{1}{\sqrt{\cosh \alpha_{nm}^{(+D)} - \cosh \left[ z + \frac{2\pi r_{-D}(m-n)}{l} \right] }}\label{PDzeta}\\&-\frac{\zeta}{\sqrt{\cosh\alpha_{nm}^{(- D)} - \cosh \left[ z + \frac{2\pi r_{-D}(m-n)}{l} \right]}}\Biggr]\notag
\end{align}
Similarly, by inserting the Wightman function in Eq.~\eqref{wightmanAB2} to Eq.~\eqref{LABgeneral}, and substituting the integration variables for $z=Y_A\tau-Y_B\tau'$ and $u=Y_A\tau+Y_B\tau'$, we obtain
\begin{align}
    L_{AB}^{(\zeta)} &=  \frac{\sigma}{\sqrt{Y_A^2 + Y_B^2}} \frac{1}{4\sqrt{\pi}l} \frac{e^{-\frac{(Y_A - Y_B)^2\sigma^2\Omega^2}{2(Y_A^2 + Y_B^2)}}}{\sqrt{\boldsymbol{\alpha}}\mathcal{N}}\sum_{n,m} \infint \D z \: e^{- \frac{z^2}{2(Y_A^2 + Y_B^2) \sigma^2}} e^{-\frac{i\Omega (Y_A + Y_B )z}{Y_A^2 + Y_B^2}} \label{LABzeta}\\&\left[\frac{1}{\sqrt{\cosh\alpha_{nm}^{(+X)} - \cosh \left( z - \beta_{nm} \right)}}-\frac{\zeta}{\sqrt{\cosh\alpha_{nm}^{(-X)} - \cosh \left( z - \beta_{nm} \right)}}\right]\notag
\end{align}
We substitute the integration variable $z$ in Eqs~\eqref{PDzeta} and \eqref{LABzeta} respectively with $\bar{z}_{nm}=z-\frac{2\pi r_{-D}}{l}(m-n)$ and $z+\beta_{nm}$
\begin{align}
P_D^{(\zeta)} &=\frac{\sigma}{4l\sqrt{2\pi}} \frac{1}{\mathcal{N}} \sum_{n,m} \int_{-\infty}^\infty d z\: e^{- \bar{z}_{nm}^2/4\sigma^2Y_D^2} e^{- i\Omega \bar{z}_{nm}/Y_D} \Biggr[\frac{1}{\sqrt{\cosh \alpha_{nm}^{(+D)} - \cosh \bar{z}_{nm} }}\label{PDzeta2}-\frac{\zeta}{\sqrt{\cosh\alpha_{nm}^{(- D)} - \cosh  z}}\Biggr]
\\
    L_{AB}^{(\zeta)} &=  \frac{\sigma}{\sqrt{Y_A^2 + Y_B^2}} \frac{1}{4\sqrt{\pi}l} \frac{e^{-\frac{(Y_A - Y_B)^2\sigma^2\Omega^2}{2(Y_A^2 + Y_B^2)}}}{\sqrt{\boldsymbol{\alpha}}\mathcal{N}}\sum_{n,m} \infint \D z \: e^{- \frac{(z+\beta_{nm})^2}{2(Y_A^2 + Y_B^2) \sigma^2}} e^{-\frac{i\Omega (Y_A + Y_B )}{Y_A^2 + Y_B^2}(z+\beta_{nm})} \label{LABzeta2}\\&\left[\frac{1}{\sqrt{\cosh\alpha_{nm}^{(+X)} - \cosh z}}-\frac{\zeta}{\sqrt{\cosh\alpha_{nm}^{(-X)} - \cosh z}}\right]\notag
\end{align}
Let us take a term from Eq.\ \eqref{PDzeta}:
\begin{align}
     &\sum_{n,m} \int_{-\infty}^\infty d z\: e^{- \bar{z}_{nm}^2/4\sigma^2Y_D^2} e^{- i\Omega \bar{z}_{nm}/Y_D} \frac{1}{\sqrt{\cosh \alpha_{nm}^{(+D)} -  \cosh z  }}
     \\
     & \qquad =\sum_{n,m} \int_{0}^\infty d z\: e^{- \bar{z}_{nm}^2/4\sigma^2Y_D^2} e^{- i\Omega \bar{z}_{nm}/Y_D} \frac{1}{\sqrt{\cosh \alpha_{nm}^{(+D)} -  \cosh z  }}\label{EqA17}
     \\ 
     & \qquad +\sum_{n,m} \int_{-\infty}^0 d z\: e^{- \bar{z}_{nm}^2/4\sigma^2Y_D^2} e^{- i\Omega \bar{z}_{nm}/Y_D} \frac{1}{\sqrt{\cosh \alpha_{nm}^{(+D)}- \cosh z }}\notag
     \intertext{For the second term in Eq.\ \eqref{EqA17}, we can substitute $z\to -z$ and flip the integration domain (with a multiple of $-1$), we can also substitute $n,m\to-n,-m$ without changing the summation domain as it sums from $-\infty$ to $\infty$. These variable and indices transformations changes $\bar{z}_{nm}=z- 2\pi r_{-D}(m-n)/l$ to $-z-2\pi r_{-D}(-m+n)/l =\bar{z}_{nm}$.}
     & \qquad =\sum_{n,m} \int_{0}^\infty d z\: e^{- \bar{z}_{nm}^2/4\sigma^2Y_D^2} e^{- i\Omega \bar{z}_{nm}/Y_D} \frac{1}{\sqrt{\cosh \alpha_{nm}^{(+D)} -  \cosh z  }}
     \nonumber 
     \\
     & \qquad -\sum_{n,m} \int_0^\infty d z\: e^{- \bar{z}_{nm}^2/4\sigma^2Y_D^2} e^{ i\Omega \bar{z}_{nm}/Y_D} \frac{1}{\sqrt{\cosh \alpha_{nm}^{(+D)}- \cosh z }}
     \label{EqA20}
     \intertext{Note that the denominators in the first and second term of Eq.\ \eqref{EqA20} are the exact same as $\cosh(\cdot)$ is an even function. Thus, we can simplify it to}\notag\\
     & \qquad =\sum_{n,m} \int_0^\infty d z\:  e^{- \bar{z}_{nm}^2/4\sigma^2Y_D^2}( e^{- i\Omega \bar{z}_{nm}/Y_D}-e^{ i\Omega \bar{z}_{nm}/Y_D}) \frac{1}{\sqrt{\cosh \alpha_{nm}^{(+D)} - \cosh z }}\notag\\
     &\qquad =2\sum_{n,m}\text{Re} \int_0^\infty d z\:  e^{- \bar{z}_{nm}^2/4\sigma^2Y_D^2} e^{- i\Omega \bar{z}_{nm}/Y_D}\frac{1}{\sqrt{\cosh \alpha_{nm}^{(+D)} - \cosh z }}
\end{align}
We can generalize this result to all terms in Eqs~\eqref{PDzeta} and \eqref{LABzeta} and obtain
\begin{align}
P_D^{(\zeta)}  &=\frac{\sigma}{2l\sqrt{2\pi}} \frac{1}{\mathcal{N}} \sum_{n,m} \text{Re}\int_{0}^\infty d z\: e^{- \bar{z}_{nm}^2/4\sigma^2Y_D^2} e^{- i\Omega \bar{z}_{nm}/Y_D} \Biggr[\frac{1}{\sqrt{\cosh \alpha_{nm}^{(+D)} - \cosh z }}
\nonumber 
\\
& \qquad -\frac{\zeta}{\sqrt{\cosh\alpha_{nm}^{(- D)} - \cosh z}}\Biggr]
\label{PDzeta2}
\\
    L_{AB}^{(\zeta)} &=  \frac{\sigma}{\sqrt{Y_A^2 + Y_B^2}} \frac{1}{2\sqrt{\pi}l} \frac{e^{-\frac{(Y_A - Y_B)^2\sigma^2\Omega^2}{2(Y_A^2 + Y_B^2)}}}{\sqrt{\boldsymbol{\alpha}}\mathcal{N}}\sum_{n,m}\text{Re} \int_0^\infty \D z \: e^{- \frac{z^2}{2(Y_A^2 + Y_B^2) \sigma^2}} e^{-\frac{i\Omega (Y_A + Y_B )z}{Y_A^2 + Y_B^2}} \nonumber 
    \\
    & \qquad \left[\frac{1}{\sqrt{\cosh\alpha_{nm}^{(+X)} - \cosh \left( z - \beta_{nm} \right)}}-\frac{\zeta}{\sqrt{\cosh\alpha_{nm}^{(-X)} - \cosh \left( z - \beta_{nm} \right)}}\right]
    \label{LABzeta2}
\end{align}

\section{Domain of $\mathcal{R}$ for Nonzero Angular Momentum}\label{Appendix:DomainR}

\noindent In this Appendix we will find the domain of $\mathcal{R}$ in Sec.\ \ref{sec:Equal angular momentum}, where the angular momentum of both black holes are set to be equal while all parameters except the detector radius and mass of black $B$ are fixed. By Eq.\ (\ref{eq:B1}), 
\begin{align}
    r_{+A} &=\frac{Jl}{2r_{-A}} \label{eq:B1}
\\
    r_{+B} &\leq R
\end{align} 
which implies,
\begin{align} 
    \mathcal{R} =\frac{r_{+B}}{r_{+A}}&\leq \frac{2R}{Jl}r_{-A} 
\end{align}
by Eq.\ \eqref{eq:B1}. Now, by Eq.\ \eqref{explicitrplus}
\begin{align}
r_{+B}&=\frac{\sqrt{l^2M_B+l\sqrt{l^2M_B^2-J^2}}}{\sqrt{2}} \geq \min_{M_B\geq 0}\left(\frac{\sqrt{l^2M_B+l\sqrt{l^2M_B^2-J^2}}}{\sqrt{2}}\right)\label{eq:B5}
\end{align}
As Eq.\ \eqref{eq:B5} is a monotonically increasing function of $M_B$, we only need to pick the lowest possible value of $M_B$, which is $J/l$
\begin{align}
    r_{+B}&\geq \sqrt{\frac{Jl}{2}}\label{eq:B6}\\
    \mathcal{R}=\frac{r_{+B}}{r_{+A}}&\geq \sqrt{\frac{2}{Jl}}r_{-A}
    \label{eq:B7}
\end{align}
by Eq.\ \eqref{eq:B1} and \eqref{eq:B6}. By \eqref{eq:B1} and \eqref{eq:B7}, we obtain
\begin{equation}
     \sqrt{\frac{2}{Jl}}r_{-A}\leq \mathcal{R}\leq \frac{2R}{Jl}r_{-A}
\end{equation}
as desired.

\section{Convergence of Image Sums in Transition Probability}\label{Appendix:nmax} 

\noindent To show that the transition probability can be appropritately expressed as an infinite sum, we compute the double sum terms in Eqs~\eqref{PDzeta2}, \eqref{LABzeta2} $n,m$ up to some integer value $n_{\text{max}}$. By plotting the resulting values at different $n_{\text{max}}$ and observe weather it approaches a certain value, we can infer the presence of convergence.

\begin{figure*}[h]
    \centering
    \includegraphics[width=\linewidth]{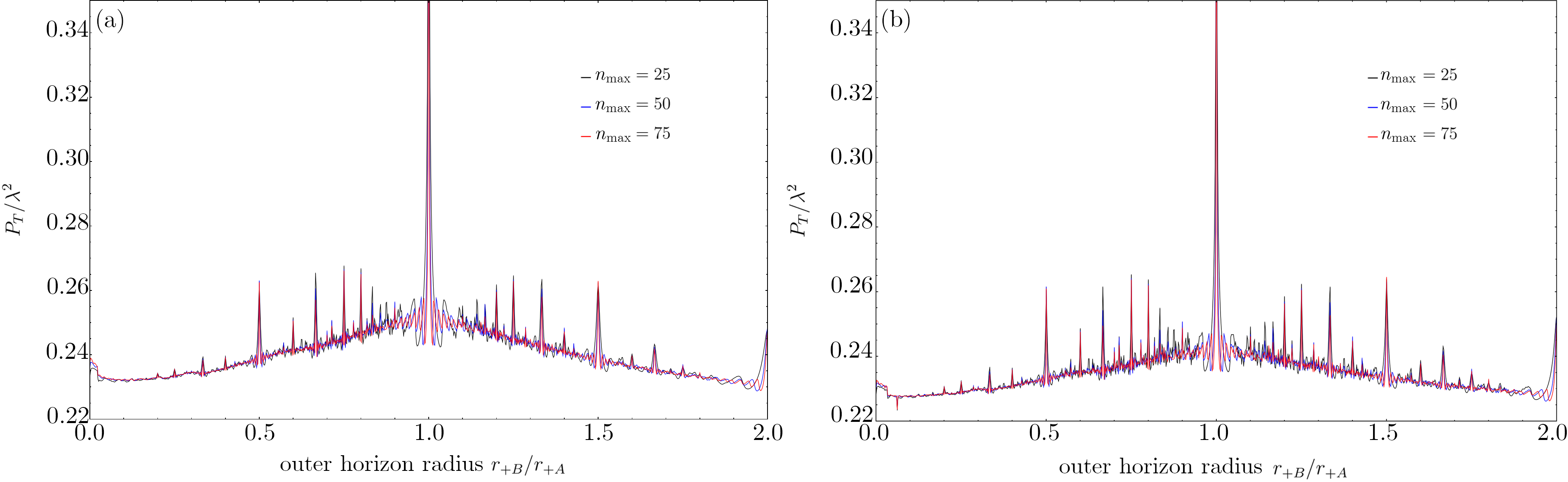}
    \caption{Transition probability as a function of $\mathcal{R}$ with parameters $\sigma \Omega = 0.02, M_A = 1, r_{-A}/l = 0.25$ and (a) $R/l = 2$, (b) $R/l = 3$ with varied $n_\mathrm{max}$}
    \label{fig:variednmax}
\end{figure*}

As we can see in \autoref{fig:variednmax}, transition probability converges to some value as $n_{max}$ increases. While the values of $\mathcal{R}$ where the prominent peaks occur converges, the number of peaks of lesser amplitudes increases while the corresponding $\mathcal{R}$ fluctuates.

\end{widetext}

\bibliography{main}

\end{document}